\documentclass[sigconf]{acmart}
\usepackage{booktabs} 

\usepackage[english]{babel}
\usepackage{moresize}
\usepackage{amsmath}
\usepackage{algorithmic}
\usepackage{balance}
\usepackage{comment}
\usepackage{paralist}
\usepackage{bm}
\usepackage{pgfplots}
\usetikzlibrary{pgfplots.dateplot}

\usepackage{flushend}
\usepackage[english]{babel}
\usepackage[latin1]{inputenc}
\usepackage{mathrsfs}
\usepackage{graphicx}

\usepackage{amssymb}
\usepackage{amsfonts}
\usepackage{url}
\usepackage{longtable}
\usepackage{rotating}
\usepackage{multirow}
\usepackage{mathrsfs}
\usepackage{subfigure}
\usepackage{enumitem}
\usepackage[linesnumbered,algoruled,boxed,lined]{algorithm2e}
\usepackage{adjustbox}
\usepackage{hyperref}
\usepackage{pgfplots}
\usetikzlibrary{pgfplots.dateplot}
\usepackage{filecontents}
\definecolor{tblue}{RGB}{31,119,180}
\definecolor{torange}{RGB}{255,127,14}
\definecolor{tgreen}{RGB}{44,160,44}
\definecolor{tred}{RGB}{214,39,40}
\definecolor{tpurple}{RGB}{148,103,189}

\newcommand{\hide}[1]{} 

\newcommand{\ie}{\textit{i}.\textit{e}.}
\newcommand{\eg}{\textit{e}.\textit{g}.} 
\newcommand{\wrt}{\textit{w}.\textit{r}.\textit{t}}

\usepackage{stmaryrd}

\def\model{MAERec}

\setcopyright{none}

\copyrightyear{2023}
\acmYear{2023}
\setcopyright{acmlicensed}\acmConference[SIGIR'23]{Proceedings of the 46th International ACM SIGIR Conference on Research and Development in Information Retrieval}{July 23--27, 2023}{Taipei, Taiwan}
\acmBooktitle{Proceedings of the 46th International ACM SIGIR Conference on Research and Development in Information Retrieval (SIGIR'23), July 23--27, 2023, Taipei, Taiwan}
\acmPrice{15.00}
\acmDOI{10.1145/3539618.3591692}
\acmISBN{978-1-4503-9408-6/23/07}

\begin{document}



\title{Graph Masked Autoencoder for Sequential Recommendation}

\author{Yaowen Ye}
\email{elwin@connect.hku.hk}
\orcid{0009-0006-7227-2306}
\affiliation{%
  \institution{University of Hong Kong}
  \city{Hong Kong SAR}
  \country{China}
}

\author{Lianghao Xia}
\email{aka_xia@foxmail.com}
\affiliation{%
  \institution{University of Hong Kong}
  \city{Hong Kong SAR}
  \country{China}
}

\author{Chao Huang}
\authornote{Chao Huang is the corresponding author.}
\email{chaohuang75@gmail.com}
\affiliation{%
  \institution{University of Hong Kong}
  \city{Hong Kong SAR}
  \country{China}
}


\begin{abstract}
While some powerful neural network architectures (\eg, Transformer, Graph Neural Networks) have achieved improved performance in sequential recommendation with high-order item dependency modeling, they may suffer from poor representation capability in label scarcity scenarios. To address the issue of insufficient labels, Contrastive Learning (CL) has attracted much attention in recent methods to perform data augmentation through embedding contrasting for self-supervision. However, due to the hand-crafted property of their contrastive view generation strategies, existing CL-enhanced models i) can hardly yield consistent performance on diverse sequential recommendation tasks; ii) may not be immune to user behavior data noise. In light of this, we propose a simple yet effective Graph \underline{M}asked \underline{A}uto\underline{E}ncoder-enhanced sequential \underline{Rec}ommender system (\model) that adaptively and dynamically distills global item transitional information for self-supervised augmentation. It naturally avoids the above issue of heavy reliance on constructing high-quality embedding contrastive views. Instead, an adaptive data reconstruction paradigm is designed to be integrated with the long-range item dependency modeling, for informative augmentation in sequential recommendation. Extensive experiments demonstrate that our method significantly outperforms state-of-the-art baseline models and can learn more accurate representations against data noise and sparsity. Our implemented model code is available at \href{https://github.com/HKUDS/GMRec}{https://github.com/HKUDS/MAERec}.

\end{abstract}




\maketitle

\section{Introduction}
\label{sec:intro}

Sequential recommendation aims to learn effective representations of users' interests and suggest future items that may be of interest to different users \cite{ye2020time,zhang2021causerec,yang2022multi}. This task has attracted considerable attention, given that user preference is time-evolving in nature in real-life, such as e-commerce~\cite{wang2020time} and video streaming~\cite{wei2019mmgcn} sites. To capture high-order transition relationships between items, extensive research efforts~\cite{wu2019session,wang2020global,chen2020handling,ma2020memory} have been devoted to proposing various graph neural networks for improving recommendation by recursively propagating information among adjacent items.

Behavior sequences of users in sequential recommender systems generally follow a long-tail distribution, in which a larger number of users merely interact with very few items~\cite{liu2020long,kim2019sequential}. While Graph Neural Networks (GNNs) have achieved promising results for approaching the sequential recommendation task in a fully supervised manner, the label data sparsity issue could significantly degrade the model representation performance~\cite{zhu2021graph,xie2020contrastive}. To tackle the label insufficiency issue, recent attempts \cite{xie2020contrastive,qiu2022contrastive,liu2021contrastive} have been made to bring the benefits of Contrastive Learning (CL) into sequential recommender systems to provide auxiliary self-supervision as augmentation signals. Following the mutual information maximization (InfoMax) framework, the key idea of these CL approaches is to reach an agreement between the embeddings encoded from two augmented views based on the InfoNCE objective. They introduce various augmentation schemes to corrupt item sequence structures based on different heuristics, \eg, stochastic item masking and reordering in CL4SRec~\cite{xie2020contrastive}, dropout-based semantic preservation in DouRec~\cite{qiu2022contrastive}, and substitute and insert operations in CoSeRec~\cite{liu2021contrastive}.

Despite recent advances, existing CL-based sequential recommender systems are severely limited by the following factors: \\\vspace{-0.12in}

\noindent (1) \textbf{Hand-crafted Contrastive Augmentation}. State-of-the-art Contrastive Learning (CL) methods still heavily rely on manually designed data corruption schemes based on heuristics to construct views for embedding contrast. Then, the representation consistency is maximized within the positive pair and minimized between negative instances~\cite{xie2020contrastive}. Guided by the inherent design of the InfoMax principle, the success of current contrastive self-supervision largely relies on high-quality augmentation with accurate contrastive view generation. However, it often requires domain knowledge with arduous labor, which can hardly be adaptable to diverse sequential recommendation scenarios with limited supervision. Furthermore, blindly corrupting the sequence structures for generating augmented views may damage the important transition structures and impair the representation learning over short sequences. \\\vspace{-0.12in}

\noindent (2) \textbf{Noise Perturbation for Data Augmentation}. Data noise issues, such as interactive behavior noise and popularity bias, commonly exist in real-life recommender systems~\cite{chen2020bias,tian2022learning,xia2022self}. Trained Self-Supervised Learning (SSL) models can be easily misled by spurious item transitional correlations, leading to suboptimal performance. Particularly, the noisy information can propagate through the item-wise connections, which negatively affects the representation learning with contrastive augmentation techniques. In the face of data noise issues, contrastive self-supervised learners are likely to overfit to the noisy labels and result in poor performance. Therefore, to build SSL-enhanced sequential recommenders that are robust to noise perturbation in practical scenarios, we need to mitigate the noisy effects in the self-supervision process with adaptive and selective augmentation for encoding interaction dynamics. \\\vspace{-0.13in}

\begin{figure}[t]
\centering
\includegraphics[width=1.0\linewidth]{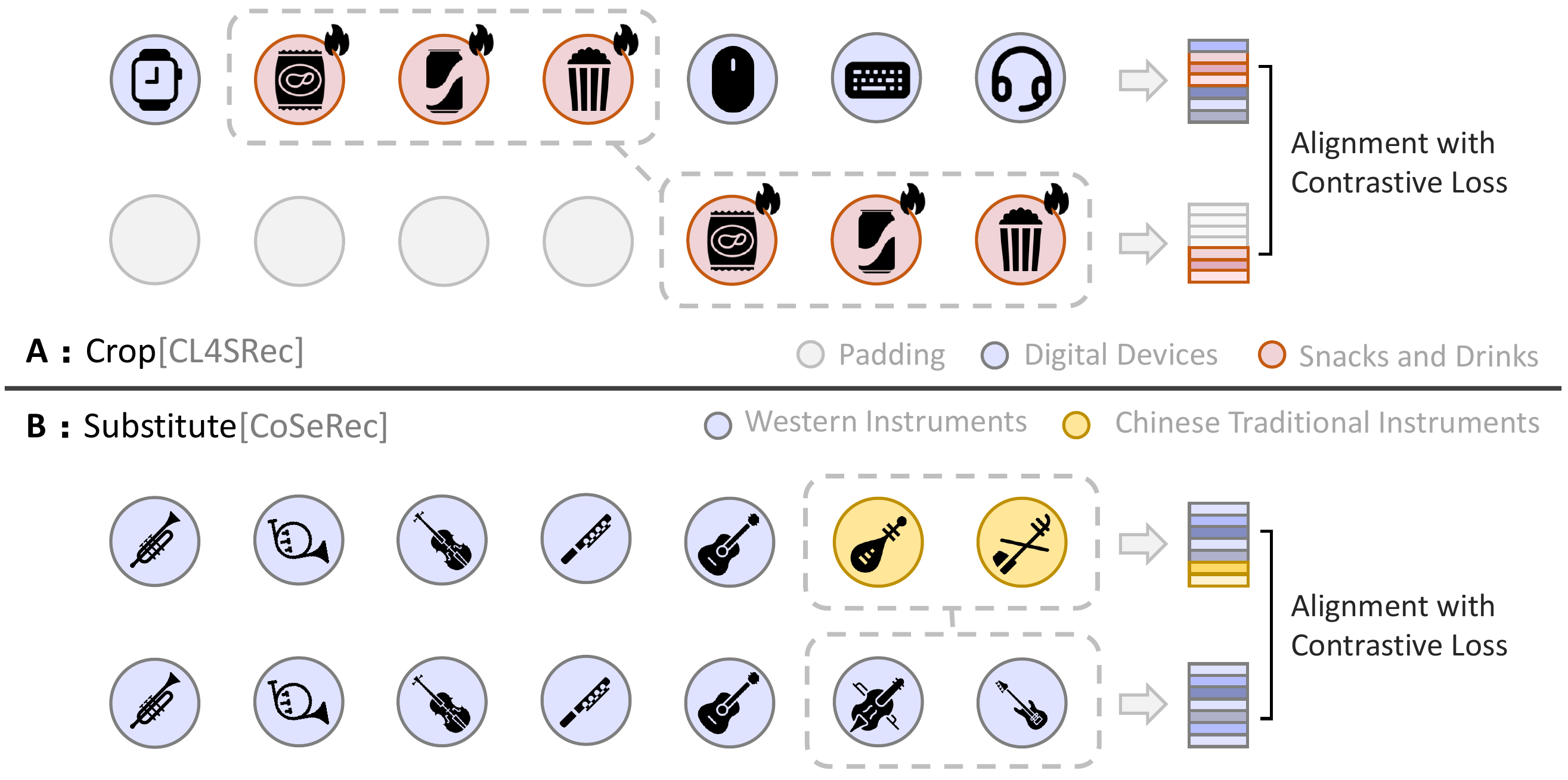}
\vspace{-0.25in}
\caption{Motivating examples of noisy and biased data augmentation in existing CL-based sequential recommenders.}
\label{fig:intro_case}
\vspace{-0.25in}
\end{figure}

Taking the augmentation strategies used in existing CL-based sequential recommenders as examples, as depicted in Figure~\ref{fig:intro_case}, user A is a digital enthusiast with the purchase of many digital products. Influenced by the New Year sale in online retailers, he/she also buys some popular snacks and drinks due to the position bias of items. However, the crop operations in CL4SRec~\cite{xie2020contrastive} may drop important interaction data and keep biased information for augmentation. In such cases, contrastive alignment between noisy augmented views will unavoidably impair the self-supervision quality and result in misleading user preference modeling. Similarly, the substitute augmentation (in CoSeRec~\cite{liu2021contrastive}) on the item sequence of user B may lose the limited but important interaction data on long-tail items (\eg, \emph{erhu}, \emph{pipa}). Some other more popular instruments (\eg, \emph{guitar}) are used for substitution, which may worsen the recommendation performance on long-tail items with limited interaction labels.

In light of the aforementioned limitations, an interesting question naturally arises: \emph{How to build SSL-enhanced sequential recommender systems that are easily adaptable and noise-resistant?} Inspired by the recent success of autoencoder-based masking techniques in image data augmentation~\cite{he2022masked,he2022vlmae}, generative self-supervised learning with the goal of masked data reconstruction can naturally alleviate the heavy reliance on manually constructing high-quality contrastive views for accurate embedding alignment.\\\vspace{-0.12in}

\noindent \textbf{Contribution}. In this work, we develop a new way to enhance sequential recommender systems with robust and adaptable self-supervised learning. To do so, we propose a lightweight and principled graph masked autoencoder model (named \model) to automate the self-supervised augmentation process. \model\ adaptively and dynamically distills informative signals for reconstruction in response to changing sequential recommendation environments.

\emph{First}, to mitigate the noisy effects during augmentation, we automatically identify items with higher semantic consistency through a learnable masking scheme. We introduce task-adaptive regularization to enhance self-supervision with an awareness of downstream recommendation tasks. \emph{Second}, we selectively mask out item transition paths with higher helpfulness and feed them into the graph autoencoder for reconstruction. By doing so, adaptive augmentation is seamlessly integrated with the modeling of sequential item dependency to cancel out the bias introduced by interaction noise and task-irrelevant information. \emph{Finally}, in accordance with adaptive augmentation by \model, we supplement the main recommendation task with noise-resistant SSL signals.

In this way, our \model\ model can not only conduct strong augmentation to train robust sequential recommenders but also regulate self-supervised learning with task-adaptive semantics via the generative reconstruction task. Our model requires no heuristics and is generalizable to various sequential interaction patterns on different datasets. Extensive experiments under various settings demonstrate the superiority of our algorithm compared to various baselines. The performed model ablation and robustness studies justify how adaptive SSL augmentation mitigates the impact of data noise and sparsity. In addition, our method achieves comparable efficiency when competing with baselines.
\section{Preliminaries and Related Work}
\label{sec:relate}

\noindent \textbf{Sequential Recommendation}.
Suppose $\mathcal{U}$ and $\mathcal{V}$ represent the sets of users and items, respectively. Each user ($u \in \mathcal{U}$) is associated with a temporally ordered sequence of their historical interacted items, denoted as \(S^u=(s^u_1, s^u_2, \cdots, s^u_{l_u})\). \(s^u_t \in \mathcal{V}\) denotes the $t$-th interacted item of user $u$, and $l_u$ is the length of $u$'s item sequence $S^u$. Given an interaction sequence $S^u$ of a user, the target of sequential recommendation is to predict the next item $s_{l_u+1}^u$ that user \(u\) is most likely to interact with (\eg, click or purchase).

To study the sequential recommendation problem, a surge of approaches have been developed to encode sequential behavioral patterns using different techniques, \eg, recurrent neural networks (RNNs)~\cite{hidasi2015session}, convolutional nerual networks~\cite{tang2018personalized,yan2019cosrec}. Inspired by the Transformer architecture, the self-attention mechanism has been adopted to deal with pairwise item correlations in user behavior sequences~\cite{kang2018self,fan2021lighter,fan2022sequential}. In addition, with the development of graph neural networks (GNNs), graph-augmented representation encoders are built upon GNN-based frameworks to capture the long-term sequential preference of users via multi-layer message passing, such as SRGNN~\cite{wu2019session}, GCE-GNN~\cite{wang2020global}, and SURGE~\cite{chang2021sequential}. However, most existing supervised approaches face the label sparsity issue, which significantly limits the model representation ability in real-life recommender systems. \\\vspace{-0.12in}

\noindent \textbf{Contrastive Learning (CL) for Sequential Recommenders}.
Recently, CL has attracted significant attention for addressing the challenge of label supervision deficiency~\cite{cailightgcl} and learning accurate representations of item sequential dependencies~\cite{lin2022dual}. In general, sequential recommendation models with contrastive learning aim to explore self-supervised signals from unlabeled data~\cite{zhang2022enhancing,yang2023debiased}. For data augmentation, existing CL-based recommender systems excessively rely on handcrafted designs to construct contrastive representation views for reaching embedding agreement.

Data corruption is performed on user behavior sequences with various augmentation operations to generate contrastive views. For example, CL4SRec~\cite{xie2020contrastive} performs sequence randomly masking and reordering operations to establish augmented contrastive views. DuoRec~\cite{qiu2022contrastive} applies model-level augmentation with dropout masks. To capture diverse user intents of time-ordered interaction behaviors, ICLRec~\cite{chen2022intent} proposes to model latent variables corresponding to user intents with contrastive self-supervision. The effectiveness of their contrastive learning methods largely depends on the high quality of generated positive and negative samples, which is not always guaranteed across different recommendation scenarios. \\\vspace{-0.12in}

\noindent \textbf{Masked Autoencoder for Representation Learning}. Inspired by the success of generative self-supervised learning in language representation~\cite{devlin2018bert}, masked autoencoders (MAE) have been proposed as effective learners for understanding images~\cite{he2022masked}. In these MAE-based vision learners, pixel-level image reconstruction is conducted using the encoder-decoder framework. Additionally, vision-language masked autoencoders have been developed to learn multi-modal embeddings to pair image-text instances~\cite{he2022vlmae}. Motivated by this masking-then-prediction paradigm~\cite{chen2022sdae}, MAE has been introduced to graph representation learning in recent attempts. For instance, GraphMAE~\cite{hou2022graphmae} focuses on reconstructing node features through randomly masking based on the scaled cosine error measurement. S2GAE~\cite{tan2022mgae} randomly drops node-wise connections from the input graph structure for reconstruction. Building upon this line, our approach incorporates the benefits of masked autoencoding and advances it with adaptive graph masking to provide meaningful self-supervisory signals for sequential recommendation.

\section{Methodology}
\label{sec:solution}

\begin{figure*}
    \centering
    \includegraphics[width=\textwidth]{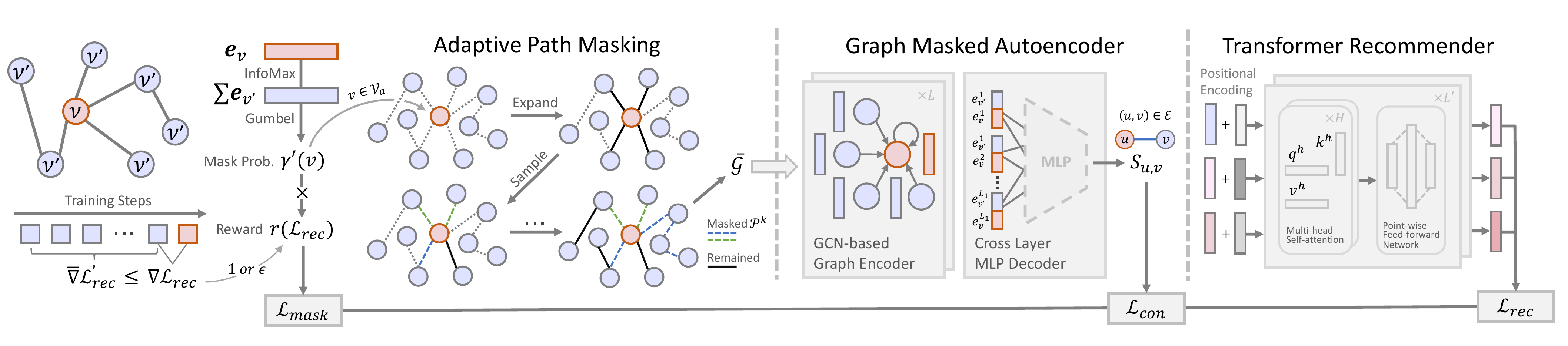}
    \vspace{-0.35in}
    \caption{\mbox{Overall architecture of our proposed \model. i) Adaptive path masking for automatically and dynamically distilling} information transition patterns for self-supervision. ii) Reconstruction of masked item transitions is performed with a \mbox{lightweight graph masked autoencoder. iii) Transformer as our backbone for sequence encoding in main recommendation task.}}
    \vspace{-0.05in}
    \label{fig:model}
\end{figure*}

In this section, we introduce our \model\ recommender system and illustrate the model architecture overview in Figure~\ref{fig:model}.

\vspace{-0.05in}
\subsection{Adaptive Transition Path Masking}
Different from the random masking strategy in some existing masked autoencoders~\cite{hou2022graphmae,tan2022mgae}, our graph masking mechanism is designed to achieve the goal of adaptive self-supervised augmentation. In this way, the graph masked autoencoder is generalized to automatically and dynamically distill useful information for masking on diverse recommendation data. Furthermore, to empower our model to preserve the global context of long-range item transitions, we propose masking graph path information with high-order item connections, instead of masking single nodes or edges.

To supercharge our model with cross-sequence item transition information and unleash the potential of graph masked autoencoder, we construct a global item-item transition graph over sequences. \\\vspace{-0.12in}

\noindent\textbf{Global Item Transition Graph}. To capture the dependence among different users, we generate a graph $\mathcal{G}=(\mathcal{V}, \mathcal{E})$ to represent the transition relationships among different items by considering user behavior sequences (\eg, $S^u$, $u\in \mathcal{U}$). The item set $\mathcal{V}$ serves as the vertex set of the transition graph. For the edge set $\mathcal{E}$, we go through all user sequences and build an edge between each item and its $h$-hop neighbors in each sequence. We count duplicated edges only once. Formally, the edge set of the transition graph is built as:
\begin{align}
    \mathcal{E} = \{(s_t^u,s_{t'}^u) : u\in\mathcal{U}, |t-t'|\leq h, 1\leq t, t'\leq l_u\}
\end{align}

\noindent Given the constructed item transition graph $\mathcal{G}=(\mathcal{V}, \mathcal{E})$, the overall adaptive path masking paradigm with learnable data augmentation policy consists of several key modules: (i) \textbf{Learning to Mask}--automatically discovering the set of anchor nodes for masked path construction; (ii) \textbf{Transition Path Masking}--masking paths underlying the graph structures to inherit informative transition patterns; (iii) \textbf{Task-Adaptive Augmentation}--recalibrating the self-supervised graph masking process with task-adaptive contexts.

\subsubsection{\bf {Learning to Mask}} The target of the learning to mask module is to learn a strategy for generating the set of anchor nodes \(\mathcal{V}_{a}\) for the learnable graph augmenter. Inspired by the InfoMax principle between path-level and graph-level embeddings~\cite{velickovic2019deep,jing2021hdmi}, we propose to measure the semantic relatedness between items by deriving their representation consistency over the graph structure of global item transitions $\mathcal{G}=(\mathcal{V}, \mathcal{E})$. In particular, we sample the \(k\)-hop transition neighbors of the target item node \(v\) on graph $\mathcal{G}$ to generate the transition subgraph for \(v\) and represent its surrounding context given the generated global item transition relationships. \\\vspace{-0.12in}

Items that have noisy interactions or are biased towards popularity may have diverse distributions in their embeddings (encoded with spurious graph connections), which can lead to suppression of their semantic relatedness scores. Therefore, we sample items with higher semantic relatedness scores as anchor nodes \(\mathcal{V}_{a}\) from $\mathcal{G}$.

To formalize this approach, we define the semantic relatedness \(\gamma(v)\) between a node \(v \in \mathcal{V}\) and its \(k^{th}\)-order subgraph as follows:
\begin{equation}
    \gamma(v) = \frac{1}{\lvert \mathcal{N}_v^k \rvert}\sum_{v'\in\mathcal{N}_v^k} \frac{\mathbf{e}_v^T \mathbf{e}_{v'}}{ \lVert \mathbf{e}_v \rVert \lVert \mathbf{e}_{v'} \rVert}
\end{equation}
\noindent where The set of \(k\)-hop neighbors of node \(v\) on the global item transition graph \(\mathcal{G}\) is denoted by \(\mathcal{N}_v^k\). The embeddings of items \(v\) and \(v'\) (\(v'\in\mathcal{N}_v^k\)) are denoted by \(\mathbf{e}_v\) and \(\mathbf{e}_{v'}\), respectively, where \(\mathbf{e}_v,\mathbf{e}_{v'}\in\mathbb{R}^d\). We observe that nodes with higher semantic relatedness scores have higher structural consistency in their neighborhood, implying that the path mask generated from such nodes captures potential cross-sequence item transition patterns while containing less noise. This makes them more beneficial for the SSL reconstruction task. \\\vspace{-0.12in}

\noindent \textbf{InfoMax-based SSL Adaptation}. To bolster the robustness of the learning-to-mask component for data augmentation, we incorporate Gumbel-distributed noise~\cite{jang2017categorical} into the process of determining the semantic relatedness of items. The formal representation is:
\begin{equation}
    \gamma '(v)=\gamma(v)-\log (-\log(\mu)), \,\,\,\,\mu \sim \text{Uniform}(0,1) .
\end{equation}
\noindent The anchor set $\mathcal{V}_{a}$, with size $\alpha$ ($|\mathcal{V}_{ac}|=\alpha$), is formed by ranking all nodes based on their semantic relatedness and using this ranking to determine their selection probability.\\\vspace{-0.12in}

To elevate the adaptability and learnability of the learning-to-mask component for data augmentation, we introduce self-supervised learning signals to the semantic relatedness by maximizing the InfoMax-based semantic relatedness. Formally, we have:
\begin{equation} \label{eq:loss_mask}
    \mathcal{L}_{mask}=-\sum_{v\in\mathcal{V}}\gamma '(v)
\end{equation}
\noindent We define the augmented loss objective $\mathcal{L}_{mask}$ to introduce additional self-supervision signals to the learning-to-mask paradigm. The main concept behind the optimized $\mathcal{L}_{mask}$ is to dynamically adjust the semantic relatedness derivation based on the impact of masking on downstream tasks. This automation of the learning-to-mask process enables it to adapt well to changing sequential recommendation scenarios.

\subsubsection{\bf{Transition Path Masking}} Once the anchor node set $\mathcal{V}_{a}$ has been generated using the learning-to-mask module, we propose to mask informative item-item transitional patterns with graph paths as SSL-based reconstruction signals. To achieve this, our transition path masking mechanism should possess two essential properties: (1) The reconstructed item transitional relations should not only contain intra-sequence item associations but also include inter-sequence item dependencies; (2) Diverse sequential patterns from both recent and past interactions should be preserved to reflect short-term and long-term item dependencies.

To achieve the above objectives, our transition path masking mechanism consists of two steps: \emph{Expand} and \emph{Sample}. Specifically, we determine the paths for masking based on a recursive random walk process that contains a succession of item-wise connections over the transition graph $\mathcal{G}=(\mathcal{V}, \mathcal{E})$. Given the $k$-th order semantic relatedness derivation with respect to representation consistency, our path masking mechanism is formally defined as follows:
\begin{equation}
    \mathcal{P}^{k} = 
    \begin{cases}
        \mathcal{V}_{a}, & \text{if \(k=1\)} \\
        \mathcal{P}^{k-1} \cup \varphi(\mathcal{N}(\mathcal{P}^{k-1}), p^k) & \text{otherwise}
    \end{cases}
\end{equation}
\noindent \( \mathcal{P}^k\) denotes the set of included item transitional connections. Here, we define $\mathcal{N}(\mathcal{P}^{k-1})$ to represent the set of nodes that are directly connected to a node in $\mathcal{P}^{k-1}$ through an edge in \(\mathcal{E}\). To guide our random walk process, we use a drop ratio \(0<p<1\), which is used in the sample operation $\varphi(\cdot,\cdot)$ that we perform on the elements in a set. In the expand step of our framework, we start a random walk from the identified anchor nodes $\mathcal{V}_{a}$ to recursively add connected items for sampling transition paths through the function $\varphi(\cdot,\cdot)$. In the drop step, we select the item transition relationships that connect the dropout nodes as transition paths for reconstruction via masked autoencoder. This results in a generated path mask of an anchor node containing sequences of varied length, with a maximum length of \(2k\). By doing so, our approach is capable of capturing the complex and dynamic nature of user behaviors. Furthermore, our framework diversifies the position of the anchor node in the sequence, which injects both short-term and long-term self-supervision signals.

\subsubsection{\bf{Task-Adaptive Augmentation}} To mitigate the impact of task-irrelevant information and data variance, we have enhanced our augmenter through task adaptation. This allows us to handle a wider variety of user behavior sequences. To achieve this, we introduce a task-adaptive function that guides the learning to mask paradigm, improving the model's generalization ability by masking more informative item transition paths for self-supervision. Otherwise, the reconstruction of task-irrelevant information would weaken the representation power of the self-supervised learning model. We introduce a task-adaptive function denoted by $r(\cdot)$:
\begin{equation}
    r(\mathcal{L}_{rec}) = \begin{cases}
        1 & \text{if \(\nabla \mathcal{L}_{rec} > \overline{\nabla}\mathcal{L}_{rec}'\)} \\
        \epsilon & \text{otherwise}
    \end{cases}
\end{equation}
\noindent $\mathcal{L}_{rec}$ represents the loss function of downstream tasks. $\nabla\mathcal{L}_{rec}$ denotes the difference between the current training step and the last step, while $\overline{\nabla}\mathcal{L}_{rec}'$ represents the averaged change of $\mathcal{L}_{rec}$ over the previous $\delta>1$ steps. The contribution of data-specific masking is measured by the loss benefits brought by the corresponding reconstruction SSL loss. This function enables the learnable masking paradigm to adjust its masking strategy based on the effect of the current mask on the downstream task. We use $\epsilon<1$ as a small constant. If the self-supervised reconstruction task of the current mask leads to a faster decrease rate of $\mathcal{L}_{rec}$, it corresponds to a smaller mask loss $\mathcal{L}_{mask}$. Otherwise, the mask loss will be greater, and the learning-to-mask module will be guided to be easily adaptable.

Given the estimated task-specific loss reward $r(\mathcal{L}_{rec})$, we enhance the InfoMax-based mask learning in Eq~\ref{eq:loss_mask} with task-adaptive regularization $\mathcal{L}_{mask}=-r(\mathcal{L}_{rec})\sum_{v\in\mathcal{V}}\gamma '(v)$. The overall adaptive path masking procedure can be formulated as follows:
\begin{equation}
    \overline{\mathcal{G}} = \text{Path-Mask}(\mathcal{G}, \mathcal{V}_{a}) = (\mathcal{V},\mathcal{E} \backslash \mathcal{P}^k) 
\end{equation}
\noindent where $\overline{\mathcal{G}}$ denotes the augmented graph with masked transition paths. $\mathcal{V}_{a}$ denotes the set of adaptively selected anchor nodes, and $\mathcal{P}^k$ denotes the set of edges in the paths generated from $\mathcal{V}_{a}$.

\vspace{-0.05in}
\subsection{Graph Masked Autoencoder}
After obtaining the graph $\overline{\mathcal{G}}$ through adaptive path masking, we feed it into a graph autoencoder framework to perform the reconstruction task for self-supervision. To improve the efficiency of the model, we use a lightweight encoder-decoder framework. In particular, we leverage a simplified graph convolutional network as the encoder for graph embedding, and a cross-layer MLP as the decoder for transition path reconstruction.

\subsubsection{\bf GCN-based Graph Encoder}
Inspired by the effectiveness and efficiency of the lightweight graph convolutional network proposed in~\cite{he2020lightgcn,chen2020revisiting}, we omit the heavy transformation and activation during the message passing on \(\overline{\mathcal{G}}\) as follows:
\begin{equation}\label{eq:gcnencoder}
    \mathbf{e}_v^{l+1} = \mathbf{e}_v^l + \sum_{v' \in \mathcal{N}_v} \mathbf{e}_{v'}^l;\,\,\,\,\,
    \mathbf{\Tilde{e}}_v = \sum_{l=1}^{L} \mathbf{e}_v^l
\end{equation}
\noindent where \(L\) denotes the total number of layers, and \(\mathcal{N}_v\) denotes the \(1\)-hop neighborhood of \(v\). \(\mathbf{e}_v^l,\mathbf{e}_{v'}^l \in \mathbb{R}^d\) are the embeddings for \(v,v'\in \mathcal{V}\) in the \(l\)-th layer. \(d\) denotes the embedding dimension of the latent space. In the last layer, the residual connection is added to alleviate the gradient vanishing problem~\cite{he2016deep}. In specific, we sum up the representations from all the hidden layers to obtain the final embedding \(\mathbf{\Tilde{e}}_v\) for an item \(v\). With this GCN-based graph encoder, the proposed model can encode the structural information of the masked graph into item embeddings for recommendation as well as reconstruction of the masked edges.

\subsubsection{\bf Decoder for Transition Path Reconstruction}
After encoding the item embeddings, the decoder is responsible for reconstructing the missing item transition relationships with masked paths on the augmented \(\overline{\mathcal{G}}\). To address the over-smoothing issue of GCNs, we use a cross-layer multi-layer perceptron (MLP) as the decoder. For a masked item-item edge \((v,v')\), we use the item embeddings of \(v\) and \(v'\) in each layer of the encoder to construct the edge embedding that corresponds to the item transitional patterns.
\begin{equation}
    \mathbf{e}_{v,v'}=\mathop{\big\lVert}\limits_{i,j=1}^{L} \mathbf{e}_{v}^i \odot \mathbf{e}_{v'}^j
\end{equation}
\noindent The number of layers in the encoder GCN is denoted by \(L\), while the concatenation operation is represented by \(\lVert\) and the element-wise multiplication by \(\odot\). Once the edge embedding is constructed, it is passed into the MLP to predict a label (either true or false) for the target item transition path.

\subsubsection{\bf SSL-based Reconstruction Objective}
The proposed model differs from conventional graph autoencoders in that it generates self-supervised signals by recovering the masked global item transition paths that are learned via our method. The supplementary objective of this self-supervised reconstruction task can be expressed:
\begin{equation}
    \mathcal{L}_{con} = - \sum_{(v,v')\in\mathcal{E} \backslash \mathcal{P}^k} \log \frac{\exp(s_{v,v'})}{\sum_{v'' \in \mathcal{V}}\exp(s_{v,v''})}
\end{equation}
\noindent The outputs of the MLP corresponding to \(\mathbf{e}_{v,v'}\) and \(\mathbf{e}_{v,v''}\) are denoted by \(s_{v,v'}\) and \(s_{v,v''}\). They represent the probability of \((v,v')\) being a masked item-item transition. To accelerate the optimization, negative sampling is employed to sample negative edges \((v,v'')\).

\subsection{Transformer as Sequence Encoder}

The effectiveness of Transformer in learning sequential behavior patterns in recommender systems, as demonstrated by models like SASRec~\cite{kang2018self} and Bert4Rec~\cite{sun2019bert4rec}, inspired us to use a Transformer for our main supervised sequential recommendation task.

\subsubsection{\bf Embedding Layer}
Each item \(v\in\mathcal{V}\) is assigned a learnable embedding \(\mathbf{e}_{v}\), which is used for generating the transition path mask during training and for graph encoding. In the embedding layer of our Transformer, a learnable positional embedding \(\mathbf{p}_i\in\mathbb{R}^d\) is added to the initial embedding of the item at position \(i\) of the input sequence. The initial sequence embedding of an interacted item sequence \(S^u\) with length \(l\) for user \(u\) is then obtained as follows:
\begin{equation}
\mathbf{E}_u = [(\mathbf{\Tilde{e}}_{s^u_1} + \mathbf{p}_1), (\mathbf{\Tilde{e}}_{s^u_2} + \mathbf{p}_2), \cdots (\mathbf{\Tilde{e}}_{s^u_l} + \mathbf{p}_l)]
\end{equation}
\noindent The generated item representation \(\mathbf{\Tilde{e}}_{s^u_l}\) is obtained by aggregating information across different graph layers, as presented in Eq~\ref{eq:gcnencoder}.

\subsubsection{\bf Multi-Head Self-Attention Blocks}
The core component of the Transformer architecture is the multi-head self-attentive mechanism, which can be formulated as follows:
\begin{gather}
    \mathbf{\hat{E}}_u^l = \mathop{\big{\lVert}}\limits_{h=1}^H \mathbf{A}^h\mathbf{E}_u^l\mathbf{\mathbf{W}_V^{h}};\,\,\,\,\,
    \mathbf{A}^h=\frac{(\mathbf{E}_u^l \mathbf{W}_Q^h)(\mathbf{E}_u^l \mathbf{W}_K^h)^T}{\sqrt{d/H};
    } \\
    \mathbf{E}_u^{l+1} = \text{ReLU}(\mathbf{\hat{E}}_u^l \mathbf{W}_1+\mathbf{b}_1)\mathbf{W}_2+\mathbf{b}_2;\,\,\,\,\,
    \mathbf{\Tilde{E}}_u = \sum_{l=1}^{L'} \mathbf{E}_u^{l}
\end{gather}
\noindent where \(L'\) and \(H\) denotes the total number of multi-head self-attention blocks and the number of heads, respectively. \(\mathbf{A}^h\) denotes the attention values of the \(h\)-th head. \(\mathbf{W}_Q^h, \mathbf{W}_K^h,  \mathbf{W}_V^h\in\mathbb{R}^{d\times(d/H)}\) are the \(h\)-th head projection matrices corresponding to query, key, and value in the attention mechanism. The parameters \(\mathbf{W}_1,\mathbf{W}_2\in\mathbb{R}^{d\times d}\) and \(\mathbf{b}_1,\mathbf{b}_2\in\mathbb{R}^d\) form a point-wise feed forward network. Residual connection is applied to obtain the final sequence embedding \(\mathbf{\Tilde{E}}_u\). Note that layer normalization and Dropout of attention values as well as inputs of each block are applied to enhance model performance.

\subsection{Multi-Task Learning}
To predict the probability of a user sequence interacting with item \(v\), we obtain the final embedding of the sequence as the output of the last multi-head self-attention block and compute its dot product with the target item embedding, \(\mathbf{e}_{v}\). We use the cross-entropy loss function to compute the optimized objective for the main supervised recommendation task. Similar to~\cite{kang2018self,xie2020contrastive}, we consider all sub-sequences of each user sequence \(S^u\) with length \(l_u\) as training data, \ie, \((s_1^u), (s_1^u, s_2^u), \cdots, (s_1^u,\cdots,s_{l_u-1}^u)\). Therefore, the loss function for the recommendation task, \(\mathcal{L}_{rec}\), is given by:

\begin{equation}
    \mathcal{L}_{rec} = -\sum_{u \in \mathcal{U}} \sum_{1 \leq t \leq l_{u}} \log \sigma(\mathbf{\Tilde{E}}_{u,t} \cdot \mathbf{\Tilde{e}}_{s_u^{t+1}}) + \log(1-\sigma(\mathbf{\Tilde{E}}_{u,t} \cdot \mathbf{\Tilde{e}}_{v^{-}_t}))
\end{equation}
\noindent \(\mathbf{\Tilde{E}}_{u,t}\) denotes the sequence embedding of \((s_1^u,\cdots,s_t^u)\) and \(v_t^{-} \not\in S_u\) is a randomly sampled negative item for \(t\)-th interacted item.

To enhance our recommender using SSL augmentation, we incorporate generative SSL signals from a graph masked autoencoder by conducting model training with multi-task learning.
\begin{equation}
    \mathcal{L} = \mathcal{L}_{rec} + \mathcal{L}_{mask} + \mathcal{L}_{con} + \lambda \lVert \mathbf{\Theta} \rVert^2_{\text{F}}
\end{equation}
\noindent \(\mathcal{L}_{mask}\) is the loss for InfoMax-based SSL adaptation, while \(\mathcal{L}_{con}\) denotes the SSL loss of masked item transition reconstruction. \(\mathbf{\Theta}\) denotes the model parameters, and \(\lambda\) is the weight decay coefficient.

\subsection{In-Depth Model Discussion}

To investigate how the proposed model avoids noisy data augmentation through adaptive transition path masking, we conduct a theoretical analysis on the mask loss of a specific item. Suppose that item \(v \in \mathcal{V}\) is a noisy item whose related reconstruction could negatively impact representation learning using the provided self-supervision signals. In this case, the gradient of its corresponding mask loss with respect to its embedding \(\mathbf{e}_v\) is:
\begin{equation}
    \frac{\partial \mathcal{L}_{mask}(v)}{\partial \, \mathbf{e}_v} = 
    -\frac{r(\mathcal{L}_{rec})}{|\mathcal{N}_v^k|} \sum_{v'\in\mathcal{N}_v^k} 
    \left( 
    \frac{\mathbf{e}_{v'}}{\lVert \mathbf{e}_v \rVert \lVert \mathbf{e}_{v'} \rVert} - 
    \frac{(\mathbf{e}_v^T \mathbf{e}_{v'}) \mathbf{e}_v}{\lVert \mathbf{e}_v \rVert ^ 3 \lVert \mathbf{e}_{v'} \rVert}
    \right).\nonumber
\end{equation}
We further compute the norm of the gradient as follows:
\begin{equation}
    \bigg{\lVert} \frac{\partial \mathcal{L}_{mask}(v)}{\partial \, \mathbf{e}_v} \bigg{\rVert} = 
    -\frac{r(\mathcal{L}_{rec})}{|\mathcal{N}_v^k| \lVert \mathbf{e}_v \rVert} \sum_{v'\in\mathcal{N}_v^k} 
    (1 - \mathbf{\bar{e}}_v^T \mathbf{\bar{e}}_{v'})
\end{equation}
\noindent In the above equation, \(\mathbf{\bar{e}}_v = \mathbf{e}_v / {\lVert \mathbf{e}_v \rVert}\) and \(\mathbf{\bar{e}}_{v'} = {\mathbf{e}_{v'}} / {\lVert \mathbf{e}_{v'} \rVert}\), and the term \(\mathbf{\bar{e}}_v^T \mathbf{\bar{e}}_{v'}\) represents the cosine similarity between the two items. Therefore, as the similarity score between \(v\) and its neighbor nodes decreases, the sum decreases and the norm increases. This emphasizes the gradient of items with low structural consistency, which improves the transition path masking paradigm with SSL adaptation. As a result, our model can mitigate the influence of behavior noise compared to existing contrastive learning solutions that use handcrafted and random augmentations.

\section{Evaluation}
\label{sec:eval}

\begin{table}[t]
\caption{Dataset Statistics}
\vspace{-0.15in}
\tabcolsep=0.06in
\begin{tabular}{cccccc}
\hline
Dataset      & \#users & \#items & \#inter. & ave.len. & density              \\ \hline
Books     & 93,403            & 54,756            & 506637             & 5.45               & \(9.94 \times 10^{-5}\) \\ \hline
Toys      & 116,429           & 54,784            & 478460             & 4.11               & \(7.50 \times 10^{-5}\) \\ \hline
Retailrocket     & 91,655            & 43,886            & 452546             & 4.94               & \(1.12 \times 10^{-4}\) \\ \hline
\end{tabular}
\label{tab:datasets}
\vspace{-0.15in}
\end{table}








\subsection{Experimental Settings}

\subsubsection{\bf Datasets}
We conduct experiments on real-world datasets that are widely used in the field of recommendation systems: Amazon Books, Amazon Toys, and Retailrocket. The first two datasets are obtained from the Amazon platform, and we use the version of users' interactions with items in the categories of books and toys. Another benchmark dataset--Retailrocket is collected from an e-commerce website. We follow the similar data pre-processing steps in~\cite{xie2020contrastive,wang2020global} to generate item sequence of user individual, and summarize data statistics in Table~\ref{tab:datasets}. \vspace{-0.05in}

\subsubsection{\bf Evaluation Protocols}
We adopt the commonly used leave-one-out strategy \cite{liu2021contrastive,sun2019bert4rec} for sequential recommendation evaluation. Specifically, for each user sequence, we us the last interacted item as the test item. We employ two evaluation metrics: Hit Ratio (\emph{HR@K}) and Normalized Discounted Cumulative Gain (\emph{NDCG@K})~\cite{wang2020next,wei2022contrastive} with (K=5,10,20), to evaluate the performance of all algorithms. \vspace{-0.05in}

\begin{table*}[t]
    \caption{Performance comparison of different methods on Amazon-books, Amazon-toys, and Retailrocket datasets.}
    \vspace{-0.12in}
    \centering
    \setlength{\tabcolsep}{0.4mm}
    \small
    \begin{tabular}{c|c|ccccccccccccc|c|c}
    \hline
    \textbf{Dataset}        & \textbf{Metric} & GRU4Rec & NARM   & SASRec & ContraRec & BERT4Rec & SRGNN  & GCE-GNN & HyRec  & SURGE  & CoSeRec & DuoRec & CL4SRec & ICLRec & \model\                                 & p-val      \\ \hline
    \multirow{6}{*}{Books}  & HR@5            & 0.3523  & 0.4064 & 0.3544 & 0.4312    & 0.5385   & 0.5472 & 0.4087  & 0.4550 & 0.4861 & 0.5684  & 0.5941 & 0.5089  & 0.5208 & \multicolumn{1}{l|}{~\textbf{0.6472}} & \(7e^{-9}\)  \\
                            & NDCG@5          & 0.2517  & 0.2955 & 0.2766 & 0.3366    & 0.4404   & 0.4404 & 0.3178  & 0.3454 & 0.3637 & 0.4540  & 0.4894 & 0.3969  & 0.4146 & \textbf{0.5201}                      & \(1e^{-7}\)  \\
                            & HR@10           & 0.4817  & 0.5340 & 0.4393 & 0.5304    & 0.6238   & 0.6458 & 0.5033  & 0.5619 & 0.6193 & 0.6666  & 0.6830 & 0.6144  & 0.6172 & \textbf{0.7533}                      & \(1e^{-8}\)  \\
                            & NDCG@10         & 0.2940  & 0.3396 & 0.3058 & 0.3685    & 0.4681   & 0.4723 & 0.3484  & 0.3800 & 0.4068 & 0.4858  & 0.5182 & 0.4310  & 0.4458 & \textbf{0.5545}                      & \(1e^{-7}\)  \\
                            & HR@20           & 0.6268  & 0.6725 & 0.5407 & 0.6440    & 0.7121   & 0.7503 & 0.6090  & 0.6778 & 0.7604 & 0.7622  & 0.7706 & 0.7216  & 0.7141 & \textbf{0.8543}                      & \(2e^{-7}\)  \\
                            & NDCG@20         & 0.3306  & 0.3737 & 0.3298 & 0.3966    & 0.4904   & 0.4987 & 0.3751  & 0.4092 & 0.4424 & 0.5100  & 0.5403 & 0.4581  & 0.4703 & \textbf{0.5801}                      & \(2e^{-7}\)  \\ \hline
    \multirow{6}{*}{Toys}   & HR@5            & 0.2247  & 0.2047 & 0.2454 & 0.2989    & 0.3717   & 0.3610 & 0.2597  & 0.2978 & 0.3075 & 0.3869  & 0.4412 & 0.3451  & 0.3483 & \textbf{0.4864}                      & \(4e^{-6}\)  \\
                            & NDCG@5          & 0.1543  & 0.1383 & 0.1854 & 0.2318    & 0.2885   & 0.2720 & 0.1880  & 0.2149 & 0.2231 & 0.2939  & 0.3450 & 0.2696  & 0.2655 & \textbf{0.3715}                      & \(3e^{-3}\)  \\
                            & HR@10           & 0.3262  & 0.3029 & 0.3299 & 0.3805    & 0.4665   & 0.4674 & 0.3628  & 0.4030 & 0.4210 & 0.4925  & 0.5424 & 0.4417  & 0.4416 & \textbf{0.6033}                      & \(7e^{-4}\)  \\
                            & NDCG@10         & 0.1888  & 0.1718 & 0.2155 & 0.2579    & 0.3191   & 0.3064 & 0.2211  & 0.2488 & 0.2596 & 0.3280  & 0.3777 & 0.3007  & 0.2971 & \textbf{0.4093}                      & \(1e^{-3}\)  \\
                            & HR@20           & 0.4700  & 0.5012 & 0.4333 & 0.4862    & 0.5780   & 0.5940 & 0.4952  & 0.5251 & 0.5752 & 0.6135  & 0.6583 & 0.5638  & 0.5652 & \textbf{0.7343}                      & \(8e^{-2}\)  \\
                            & NDCG@20         & 0.2240  & 0.2470 & 0.2413 & 0.2843    & 0.3472   & 0.3383 & 0.2545  & 0.2797 & 0.2984 & 0.3585  & 0.4070 & 0.3315  & 0.3271 & \textbf{0.4424}                      & \(3e^{-2}\)  \\ \hline
    \multirow{6}{*}{Retail} & HR@5            & 0.4164  & 0.5586 & 0.6136 & 0.6587    & 0.7596   & 0.6973 & 0.4238  & 0.4801 & 0.7091 & 0.8063  & 0.7873 & 0.6663  & 0.7818 & \textbf{0.8577}                      & \(2e^{-10}\) \\
                            & NDCG@5          & 0.2950  & 0.4238 & 0.5077 & 0.5553    & 0.6882   & 0.5974 & 0.3371  & 0.3792 & 0.5902 & 0.7117  & 0.7128 & 0.5752  & 0.6904 & \textbf{0.7660}                      & \(4e^{-8}\)  \\
                            & HR@10           & 0.4518  & 0.6757 & 0.6968 & 0.7387    & 0.8084   & 0.7719 & 0.5182  & 0.5722 & 0.7857 & 0.8530  & 0.8291 & 0.7336  & 0.8282 & \textbf{0.8959}                      & \(3e^{-8}\)  \\
                            & NDCG@10         & 0.3456  & 0.4600 & 0.5389 & 0.5823    & 0.7039   & 0.6216 & 0.3676  & 0.4091 & 0.6151 & 0.7269  & 0.7264 & 0.5969  & 0.7055 & \textbf{0.7784}                      & \(1e^{-7}\)  \\
                            & HR@20           & 0.6095  & 0.7879 & 0.7687 & 0.8140    & 0.8577   & 0.8425 & 0.6275  & 0.6361 & 0.8581 & 0.8951  & 0.8703 & 0.8018  & 0.8733 & \textbf{0.9218}                      & \(1e^{-4}\)  \\
                            & NDCG@20         & 0.3830  & 0.4972 & 0.5558 & 0.6009    & 0.7164   & 0.6394 & 0.3951  & 0.4256 & 0.6334 & 0.7376  & 0.7368 & 0.6142  & 0.7169 & \textbf{0.7850}                      & \(1e^{-6}\)  \\ \hline
    \end{tabular}
    \label{tab:overall_performance}
    \end{table*}

\subsubsection{\bf Compared Algorithms}
We compare \model\ with 13 competitive baseline methods for sequential recommendation covering various techniques, which can be grouped into four categories: (1) RNN-based approaches (\ie, GRU4Rec~\cite{hidasi2015session}, NARM~\cite{kang2018self}). (2) Transformer-based methods (\ie, SASRec~\cite{kang2018self}, BERT4Rec~\cite{sun2019bert4rec}). (3) GNN-enhenced sequential recommenders (\ie, SRGNN~\cite{wu2019session}, GCE-GNN~\cite{wang2020global}, HyRec~\cite{wang2020next}, and SURGE~\cite{chang2021sequential}). (4) Self-supervised learning models (\ie, ContraRec~\cite{wang2023sequential}, CL4SRec~\cite{xie2020contrastive}, CoSeRec~\cite{liu2021contrastive}, DuoRec~\cite{qiu2022contrastive}, and ICLRec~\cite{chen2022intent}). Baselines are elaborated as follows:

\begin{itemize}[leftmargin=*]

\item \textbf{GRU4Rec}~\cite{hidasi2015session}: This method uses GRU to encode user sequences and adopts a ranking based loss for model training.\\\vspace{-0.12in}

\item \textbf{NARM}~\cite{li2017neural}: It improves GRU-based sequential recommendation models with attention mechanism as hybrid sequence encoders.\\\vspace{-0.12in}

\item \textbf{SASRec}~\cite{kang2018self}. This is a seminal model that uses a uni-directional Transformer to encode user sequential patterns over items.\\\vspace{-0.12in}

\item \textbf{BERT4Rec}~\cite{sun2019bert4rec}: This model utilizes a bi-directional Transformer as backbone for sequence encoding. The self-attention is integrated with feed-forward network for item transition modeling.\\\vspace{-0.12in}

\item \textbf{SRGNN}~\cite{wu2019session}: This model uses a gated GNN to learn item embeddings over transition graph, and leverages self-attention mechanism to generate sequence embeddings for making predictions. \\\vspace{-0.12in}

\item \textbf{GCE-GNN}~\cite{wang2020global}: The proposed model includes two types of item embeddings. The global-level embeddings capture the overall relationships between items across different sequences. Local-level item embeddings encoded by graph convolutional networks.

\item \textbf{HyRec}~\cite{wang2020next}: This model utilizes hypergraph convolutional networks to learn high-order correlations between items. It treats each user as a hyperedge to connect multiple interacted items. \\\vspace{-0.12in}

\item \textbf{SURGE}~\cite{chang2021sequential}: It is a state-of-the-art model which explicitly learns user interests through clustering and constructing item-item interest graphs based on metric learning. Then, it uses GNNs with graph pooling layers to generate embeddings for prediction. \\\vspace{-0.12in}

\item \textbf{ContraRec}~\cite{wang2023sequential}: This model uses the Transformer as backbone for sequence encoding of item transitions. For data augmentation, it proposes a holistic joint-learning paradigm to learn from multiple contrastive self-supervised signals. \\\vspace{-0.12in}

\item \textbf{CL4SRec}~\cite{xie2020contrastive}: This is a contrastive method that performs random corruptions over sequences using item cropping, masking, and reordering to generate contrastive views. The method then maximizes representation consistency between the generated contrastive views with the stochastic augmentors. \\\vspace{-0.12in}

\item \textbf{CoSeRec}~\cite{wang2023sequential}: Apart from the three contrastive views proposed in CL4SRec, this model introduces two additional augmentation operators that leverage item correlations, \ie\ substitution and insertion, for generating views for contrastive learning. \\\vspace{-0.12in}

\item \textbf{DuoRec}~\cite{qiu2022contrastive}: It introduces a contrastive regularization with model-level augmentation to improve the item embedding distribution and tackle the representation degeneration problem. \\\vspace{-0.12in}

\item \textbf{ICLRec}~\cite{chen2022intent}: This model first learns latent variables to represent user intents through clustering. It then maximizes agreement between sequences and their corresponding intents.

\end{itemize}

\subsubsection{\bf Hyperparameter Settings}
We implement our proposed \model\ using PyTorch, and use the Adam optimizer for parameter inference with a learning rate of \(1e^{-3}\) and a batch size of 256. In our graph autoencoder principle, we set the number of GNN layers in the range of \{1,2,3\}. The number of Transformer layers is set to 2 with 4 heads for multi-dimensional representation. In the adaptive transition path masking module, we sample \(\mathcal{V}_{a}\) every 10 epochs to improve efficiency. The number of anchor nodes \(\alpha\) for each masking step was searched from \(\{50, 100, 200, 400\}\). The parameter \(k\) is searched from \(\{3,5,7\}\) to determine the maximum length \(2k\) of random walk paths. The weight-decay factor is chosen from \(\{1e^{-3}, 1e^{-5}, 1e^{-7}\}\) during model training process.

\subsection{Performance Comparison}
Table \ref{tab:overall_performance} presents the performance of all methods on three benchmark datasets in terms of \emph{HR@K} and \emph{NDCG@K} under top-5, top-10, and top-20 settings. Additionally, we perform a significance test to demonstrate the superiority of \model\ over the strongest baselines, where a \(p\)-value \(<0.05\) indicates statistically significant improvement. The following observations can be made:

\begin{itemize}[leftmargin=*]

\item \textbf{Obs.1: Superiority over SOTA GNN-based models}. \model\ consistently outperforms all GNN-based baselines across different settings. Existing GNN-based models may suffer from the over-fitting issue over short sequences with insufficient training labels. Additionally, SSL-enhanced sequential recommenders perform better than GNN-based baselines, indicating the helpfulness of incorporating self-supervision signals to augment sequence encoding with limited labels. To mitigate the data scarcity problem, our \model\ incorporates generative self-supervision signals with graph masked autoencoder for effective data augmentation. As a result, informative as well as robust representations can be learned to preserve global item transition patterns. \\\vspace{-0.12in}

\item \textbf{Obs.2: Superiority over SOTA SSL methods}. \model\ achieves the best overall performance in comparison to all baselines with self-supervised augmentation. While these methods generally perform better than GNN-based models by introducing self-supervision signals, their random data augmentation still leads to sub-optimal performances. This is because random augmentation strategies (\eg item cropping or reordering) used in common SSL models are performed over all user sequences and all items, making the model unstable to noise perturbations on both sequence and item levels, such as interaction noise and popularity bias. Instead of random augmentations, \model\ adopts a learnable strategy for introducing useful self-supervision signal by masking informative transition paths generated from adaptively selected items. As a result, the newly designed \model\ suppresses existing SSL baselines by a large margin with an adaptive graph masked autoencoder as a robust data augmentor. \\\vspace{-0.12in}

\item \textbf{Obs.3: Advantage of our mask-based data augmentation}. There are several baseline models that also utilize the mask operator for self-supervised learning, such as CL4SRec and CoSeRec, which mask items in a sequence to produce contrastive views. Additionally, BERT4Rec incorporates the Cloze objective into the sequence encoder for item embeddings. Benefitting from such mask-based data augmentation, BERT4Rec outperforms SASRec, which only uses a single-directional Transformer for sequence encoding. Meanwhile, CoSeRec and CL4SRec both achieve better results than many other baselines due to their data augmentation. However, our \model\ is the only model that automates the mask-and-reconstruct paradigm over item transitional patterns with effective task-adaptation, resulting in even better performance.

\end{itemize}

\vspace{-0.1in}
\subsection{Ablation Study}
This section presents the ablation study conducted on three variants of the proposed \model\ algorithm: \textbf{-L2M}, \textbf{-PA}, and \textbf{-TA}. The purpose of the study was to investigate the contribution of different components to the overall performance of the algorithm. The results of the study are reported in Table \ref{tab:ablation_study}.

\subsubsection{\bf Effect of the Mask Generator}
The first variant, \textbf{-L2M}, replaces the our designed mask generator with random masking of the same proportion. The results indicate that non-adaptive masking may harm important transition relations or introduce noisy information to the masked-based reconstruction task, which weakens the representations and leads to suboptimal performance.

\subsubsection{\bf Effect of Transition Path Masking}
The second variant, \textbf{-PA}, studies the benefits of masking transition paths instead of single nodes. In this variant, the maximum length of a masked path is set to \(k = 1\), which makes path masking equivalent to single node masking. The results demonstrate that the proposed transition path masking significantly improves the overall performance, highlighting the effectiveness of learning to mask item transition patterns from both intra-sequence and inter-sequence perspectives.

\subsubsection{\bf Effect of Task-Adaptive Augmentation}
The adaptive masking loss with the task-adaptive function \(r(\cdot)\) is also a crucial component of our \model\ algorithm. In order to study its impact, we remove the loss for learnable masking  \(\mathcal{L}_{mask}\) (Eq~\ref{eq:loss_mask}) in the variant \textbf{-TA}, disabling the task-adaptive regularization. The results show a significant drop in performance, indicating that the task-adaptive function guides the model training in a better direction according to the gradient from the target recommendation task. With the task-adaptive constraints, our \model\ algorithm can effectively prevent the data augmentor from learning task-irrelevant information by selectively applying self-supervision augmentation.

\begin{table}[t]
\small
\caption{Model ablation study with variants.}
\vspace{-0.12in}
\begin{tabular}{ccccc}
\hline
\multirow{2}{*}{Method} & \multicolumn{2}{c}{Amazon books} & \multicolumn{2}{c}{Retailrocket} \\
                                 & HR@10     & NDCG@10     & HR@10     & NDCG@10     \\ \hline\hline
-L2M                             & 0.7482             & 0.5510               & 0.8800             & 0.7547               \\
-PA                              & 0.7272             & 0.5239               & 0.8590             & 0.7056               \\
-TA                              & 0.7507             & 0.5542               & 0.8801             & 0.7564               \\\hline
\model\                             & \textbf{0.7723}     & \textbf{0.5720}     & \textbf{0.8921}     & \textbf{0.7659}     \\ \hline
\end{tabular}
\label{tab:ablation_study}
\end{table}
\vspace{-0.1in}

\begin{figure}[t]
    \centering
    \includegraphics[width=\columnwidth]{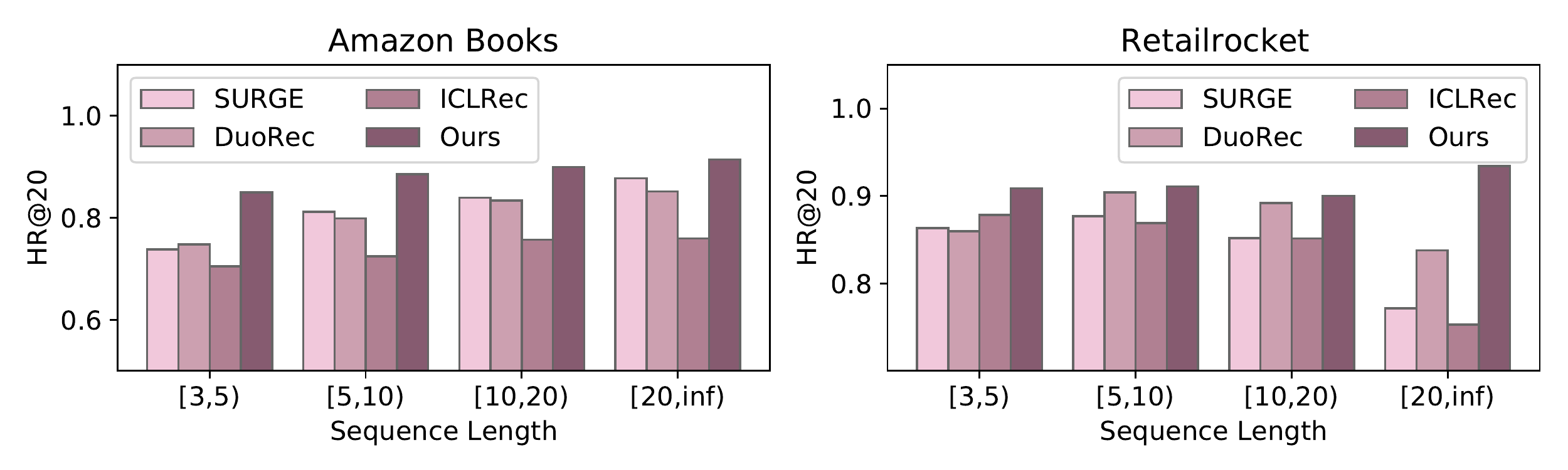}
    \vspace{-0.3in}
    \caption{Performance \wrt\ sequence length on Amazon Books and Retailrocket datasets in terms of \emph{HR@20}.}
    \vspace{-0.1in}
    \label{fig:sparsity}
\end{figure}

\begin{figure}[t]
    \centering
    \includegraphics[width=\columnwidth]{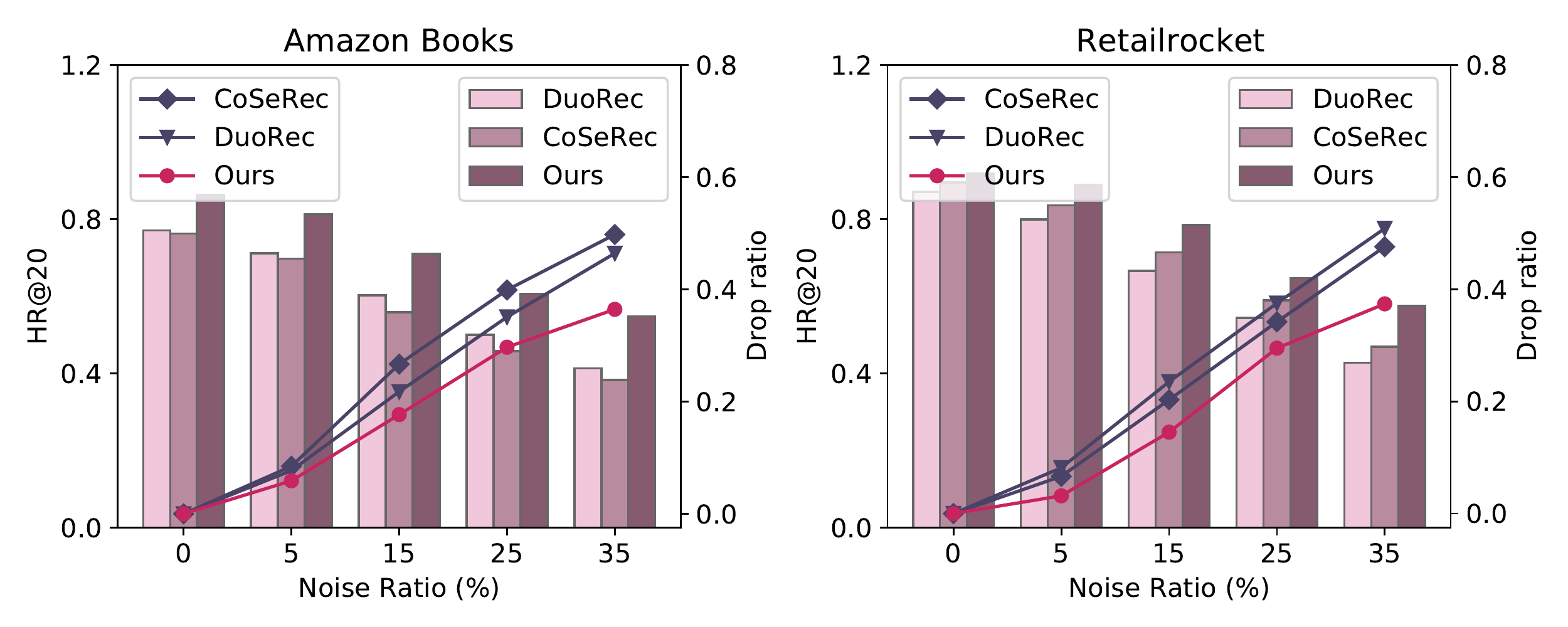}
    \vspace{-0.3in}
    \caption{Performance \wrt\ noise ratio on Amazon Books and Retailrocket. The bars show performance in terms of \emph{HR@20}, and the lines show performance degradation ratio.}
    \vspace{-0.15in}
    \label{fig:noise}
\end{figure}

\subsection{Advantage against Noise and Sparsity}
\subsubsection{\bf Performance against data sparsity}
The data sparsity issue is a common problem for sequential recommender systems with a large number of short sequences. To evaluate the robustness of our proposed \model\ algorithm against this problem, we partition the user sequences into four groups based on the length of their item sequences: \([3,5),[5,10),[10,20),[20,\infty)\). Figure \ref{fig:sparsity} shows the performance of our model and baseline methods on both Amazon Books and Retailrocket datasets. The results demonstrate that our proposed model consistently outperforms the baseline methods at all sparsity levels. Specifically, our model performs well on short sequences, which can be beneficial for addressing the label insufficiency limitation in practical recommendation scenarios. This can be attributed to the SSL-enhanced transition path masking module, which preserves both local and global item dependencies in an automatic and adaptive manner. The handcrafted augmentation used in DuoRec and ICLRec may limit their representation generality in encoding accurate user preferences from short sequences, due to the diverse nature of interaction data. In contrast, our \model\ algorithm can effectively transfer knowledge from long item sequences into the modeling of short sequences, allowing it to learn meaningful representations from sparse data with limited labels.

\subsubsection{\bf Performance against data noise}
Real-world recommendation systems often face noisy user interactions. To evaluate the robustness of our proposed \model\ algorithm against noisy data, we replace a certain proportion (\ie, 5\%, 15\%, 25\%, and 35\%) of user interactions with randomly sampled negative items, and evaluate the models on datasets with artificial noise perturbations. As shown in Figure \ref{fig:noise}, our proposed model outperforms the baselines under all noise levels on both Amazon Books and Retailrocket datasets. Moreover, our model exhibits a consistently lower performance degradation ratio compared to the baselines.

We attribute this superiority to the adaptive transition path masking module in our algorithm. Firstly, the adaptive masking strategy avoids masking edges connected to noisy nodes using the InfoMax-based mask probability and mask loss, encouraging the model to mask and learn on nodes with reliable dependency relationships. This significantly increases the model's robustness against noisy perturbations. Secondly, the transition paths in our algorithm are generated in the local subgraph of the anchor nodes, resulting in higher semantic relatedness among items in such sequences compared to randomly generated augmented sequences, such as those generated by random masking, dropping, and reordering. As a result, it can be observed in Figure \ref{fig:noise} that CoSeRec is relatively unstable against data noise, indicating the disadvantage of random augmentation methods compared to adaptive path masking.

\begin{figure}[t]
    \centering
    \vspace{-0.08in}
    \includegraphics[width=\columnwidth]{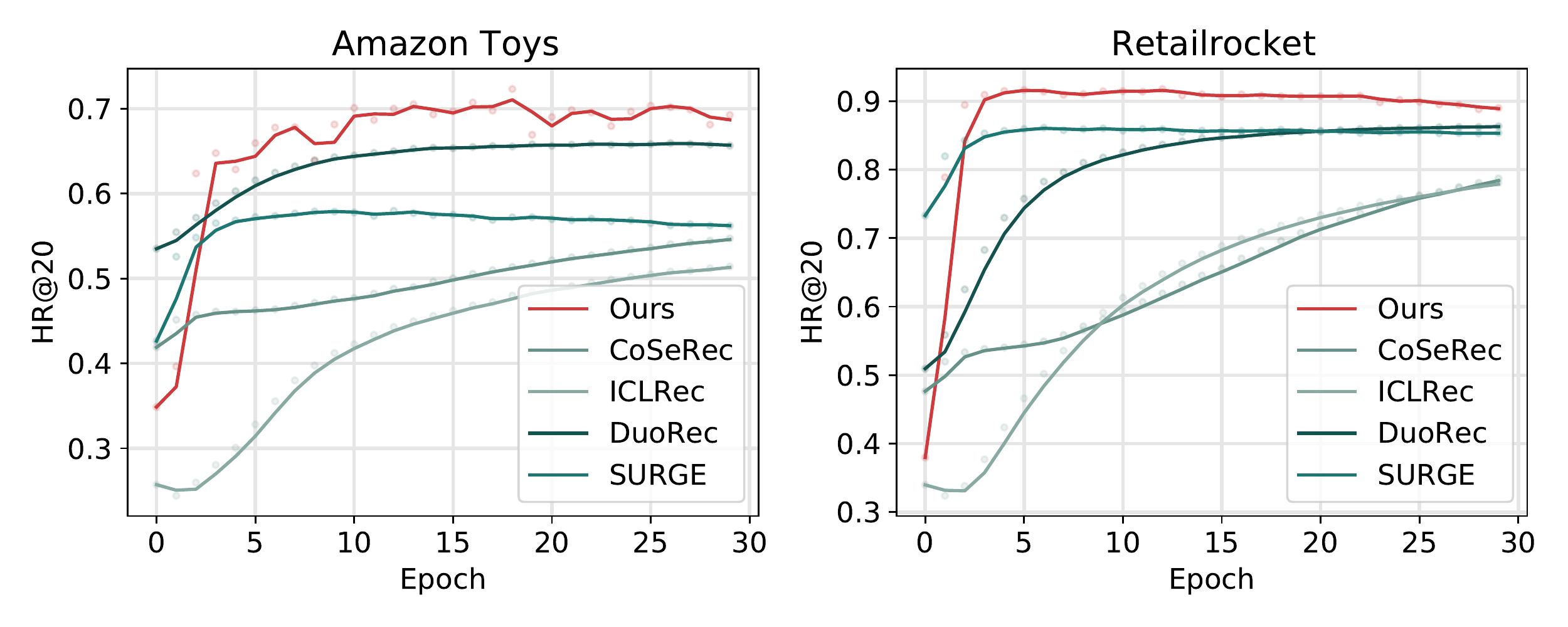}
    \vspace{-0.35in}
    \caption{Test performance \textit{w.r.t.} training epochs on Amazon Books and Retailrocket datasets for convergence analysis.}
    \vspace{-0.15in}
    \label{fig:convergence}
\end{figure}

\subsection{Model Scalability Study }
We evaluate the efficiency of our proposed model from two aspects: convergence speed and training time cost. All experiments are conducted on a single NVIDIA GeForce RTX 3090 GPU.\vspace{-0.05in}

\subsubsection{\bf Convergence Analysis}
To study the convergence efficiency of our proposed model, we compare its training process with several baselines, including both GCN-based methods (SURGE) and SSL-based methods (DuoRec, CoSeRec, ICLRec), on the Amazon Toys and Retailrocket datasets. The training process of these models is presented in Figure \ref{fig:convergence}. From the results, we observe that our proposed model achieves the best performance with the least training epochs. It reaches the best reported performance at around 10 epochs on Amazon Toys and around 5 epochs on Retailrocket, while CoSeRec takes more than 200 epochs to converge. We attribute this superiority to the self-supervised learning task enhanced by graph masked autoencoder, which injects sufficient supervision signals for the model to learn informative item transition patterns on the masked item transition graph in an efficient way. \vspace{-0.05in}

\begin{table}[t]
\caption{\mbox{Training time cost on Amazon toys and Retailrocket.}}
\vspace{-0.15in}
\begin{tabular}{cccccc}
\hline
\textbf{Data}     & SURGE & ICLRec & CoSeRec & DuoRec & Ours  \\ \hline\hline
\textbf{Amazon}  & 7.39h & 5.28h  & 21.01h & 3.89h  & 1.12h \\
\textbf{Retail} & 2.67h & 4.3h   & 16.18h  & 4.45h  & 1.82h \\ \hline
\end{tabular}
\label{tab:training_time}
\vspace{-0.2in}
\end{table}

\subsubsection{\bf Computational Cost Evaluation}
To study the training computational cost of our proposed model, we also record the training time of our model and the baselines, as shown in Table \ref{tab:training_time}. The results show that our proposed model has a significant efficiency improvement over compared methods, which suggests that our model is a lightweight and strong recommender, enabling automated and task-adaptive masked autoencoding. Our model is supercharged with adaptive SSL augmentation, and thus requires much less training time compared to other SSL-based methods, which use non-adaptive (heuristic-based) augmentation operators for downstream recommendation tasks.

\vspace{-0.05in}
\subsection{Hyperparameter Study}
\begin{figure}[t]
    \centering
    \vspace{-0.02in}
    \includegraphics[width=\columnwidth]{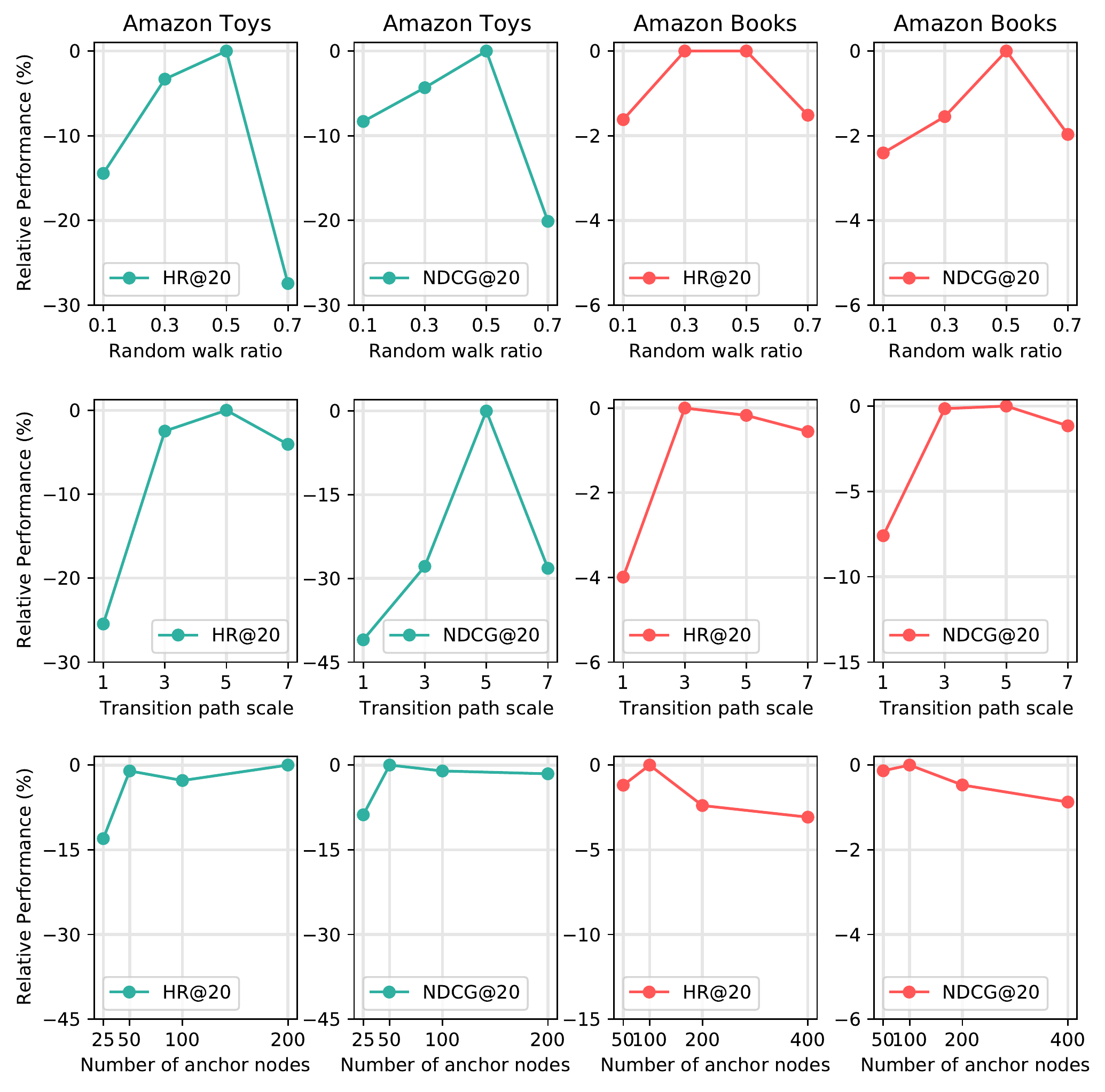}
    \vspace{-0.2in}
    \caption{Hyperparameter study of \model.}
    \vspace{-0.1in}
    \label{fig:hyp}
\end{figure}

We conduct experiments to investigate the effect of several key hyperparameters on the performance of our proposed model. The experiment results are presented in Figure \ref{fig:hyp}, in terms of the relative decrease of HR@20 and NDCG@20 on the Amazon Books and Amazon Toys datasets. The observations are shown below:

\begin{itemize}[leftmargin=*]

\item \textbf{Random walk ratio \(p\) over item transitions}. 
This hyperparameter controls the number of masked edges triggered by the identified anchor node and affects the length of masked paths, which reflect item transitional patterns. Specifically, a larger value of $p$ indicates a longer masked item transition path. Our experiments show that, $p=0.5$ achieves the best performance. This indicates that masking transition paths with variable lengths benefits the model learning with more diverse SSL signals.

\item \textbf{Masked transition path scale \(k\)}. This hyperparameter controls the maximum length $2k$ of the generated transition paths. Our experiments show that setting $k=1$ (\ie, only masking edges in the 1-hop of anchor nodes) results in significant performance degradation. Interestingly, we also observe that masking too long paths can lead to a performance drop, as insufficient information is provided for achieving accurate reconstruction results. This may impair the quality of the auxiliary SSL task for augmenting the target recommendation task.

\item \textbf{Number of anchor nodes \(\alpha\)}. 
\(\alpha\) determines the number of seed nodes for transition path masking. Experiments show that masking transition paths from too few anchor nodes leads to worse performance. On the other hand, selecting too many anchor nodes for path masking may damage the structure of the transition graph, and hence negatively affect model performance.

\end{itemize}

\begin{figure}[h]
    \centering
    \includegraphics[width=\columnwidth]{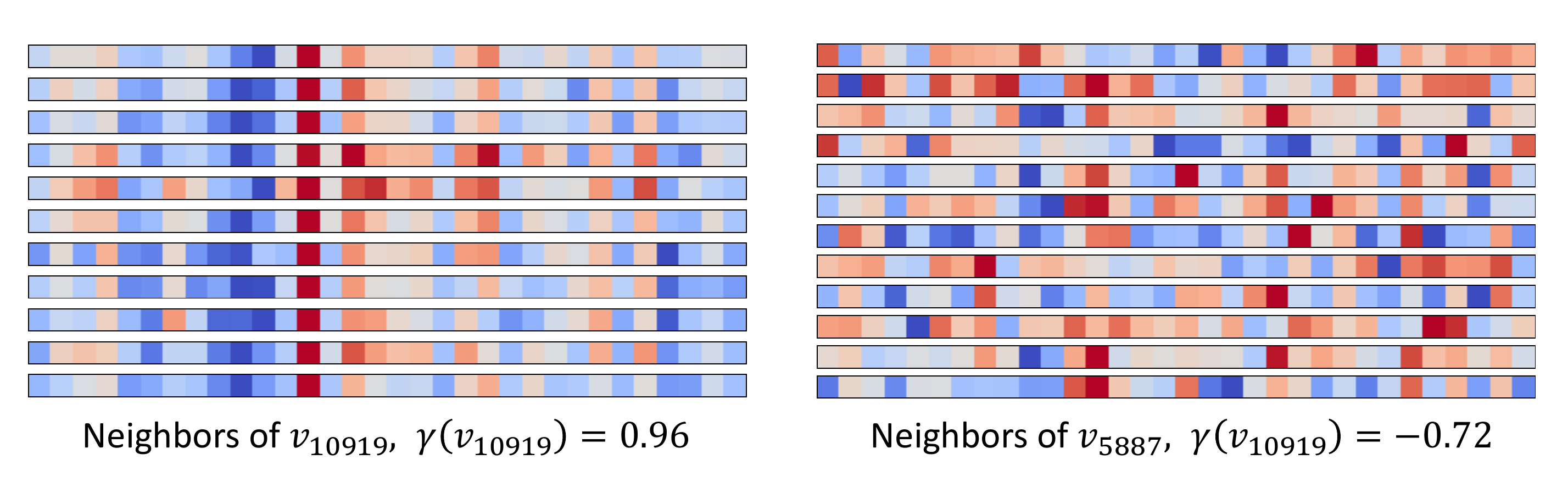}
    \vspace{-0.25in}
    \caption{Learned semantic relatedness reflects representa-\mbox{tion consistency among items connected in transition graph.}}
    \vspace{-0.1in}
    \label{fig:case1}
\end{figure}

\subsection{Case Study}
We perform a case study on the Retailrocket dataset to investigate the potential interpretation ability of our proposed model. Specifically, we select two items with high and low semantic relatedness, respectively, and visualize the embeddings of their neighbor items on the item transition graph in Figure \ref{fig:case1}. The results suggest that our learning-to-mask module is able to discover items with reliable dependency relationships to be anchor nodes, which benefits the reconstruction task. In contrast, items with low semantic relatedness (\eg, misclicked items) have inconsistent neighborhoods in terms of their embeddings. Masking transition paths from such items may result in noisy data augmentation and misguide the self-supervision process for sequential pattern encoding.

\section{Conclusion}
\label{sec:conclusoin}

This work proposes an adaptive data augmentation approach to enhance sequential recommender systems through a new graph masked autoencoder. Our proposed \model\ provides controllable and task-adaptive augmentation with strong self-supervision signals, thanks to the designed adaptive path masking. We conduct extensive experiments on three real-world datasets and demonstrat the superior performance of our \model\ compared with state-of-the-art baselines. In future work, we plan to improve the stability of SSL training by generalizing our approach to out-of-distribution sequences. This will help to address the data distribution shift between the training and test data in sequential recommendation, and generalize the model to newly arrived item sequences in future.


\clearpage
\balance
\bibliographystyle{ACM-Reference-Format}
\bibliography{sample-base}

\end{document}